%% file: main.tex
\documentclass{article}

\usepackage[preprint]{neurips_2026_vericode}
\usepackage[utf8]{inputenc}
\usepackage[T1]{fontenc}
\usepackage{hyperref}
\usepackage{url}
\usepackage{booktabs}
\usepackage{graphicx}
\usepackage{amsmath}
\usepackage{microtype}
\usepackage{xcolor}
\definecolor{linkblue}{HTML}{1A4C8B}
\hypersetup{colorlinks=true,linkcolor=black,citecolor=black,filecolor=black,
            urlcolor=linkblue,pdfborder={0 0 0}}
\newcommand{\arxivref}[1]{\href{https://arxiv.org/abs/#1}{\emph{arXiv:#1}}}
\newcommand{\doiref}[1]{\href{https://doi.org/#1}{doi:\nolinkurl{#1}}}
\input{tables/macros}

\workshoptitle{AI for Verifiable Coding}

\title{Does Fault Localization Beat a Fresh Attempt?\\
A Placebo-Controlled Study of Test-Guided Code Repair}

\author{%
  Anik Jha \\
  Independent Researcher \\
  \texttt{anik.k.jha@gmail.com} \\
}

\begin{document}
\maketitle

\begin{abstract}
Fault localization can focus a code model's repair on the statements a failing test implicates, but
a targeted edit may succeed merely because it is small, and a second model call may succeed without
using the failure at all. We separate these explanations with three arms applied to the same failed
candidate (blind whole-solution resampling, spectrum-based localization followed by suspect-span
infilling, and same-length infilling at a disjoint random code span) across three frozen 26--32B
models, three benchmarks and \DeadCandidates{} failing candidates, plus a separately declared 24B
fourth model from a third family. Three results follow. First,
localization is rarely available: only \PublicLocalizableRate{} of failing candidates expose a
failing public test with a usable spectrum, because the largest group of failures passes every test
the user can see.
Second, among the \StrongPairs{} candidates localizable from a strong suite, localized infilling
loses decisively to blind resampling at a matched number of attempts (\ResampleDiscordance{},
$p={\ResampleP}$), in the direction opposite to our hypothesis. That loss replicates in a third
family: on \MistralPairs{} candidates from Mistral-Small-24B, run on two of the three benchmarks, it
is \MistralDiff{} points (\MistralLocalizedRate{} against \MistralResampleRate{}, 95\% CI
\MistralCI{}, discordance \MistralDisc{}), the largest margin we measure, so our rule's two-family
replication clause is satisfied only in the inverted direction, by the method losing to its own
comparator. Third, giving the span arm more room does not rescue it: widening the edit until the
median span grows from \WidenSpanFrom{} line to \WidenSpanTo{} lines leaves it still losing, and an
adaptive policy adds at most \AdaptiveOverWideMax{} points over always editing wider.
Against the random-span placebo localized infilling leads
pooled (\PlaceboDiscordance{}, Holm-adjusted $p={\HolmPlacebo}$), but that lead resolves in no
individual model under the per-model attempt-level analysis our shipped analysis plan designates
primary (best Holm $p={\PrimaryPlaceboBestHolm}$), so we report the location effect as suggestive
rather than established. Re-pricing attempts as tokens narrows but does not overturn this: a span
attempt spends \SpanTokens{} generated tokens against \ResampleTokens{}, yet \SampleBudget{}
localized attempts reach \SpanCeiling{} while one blind attempt already reaches
\BlindFirstAttempt{}. The mechanism is
visible in the generations: infilling reproduces the removed span verbatim in \LocalizedNoop{} of
attempts and yields \LocalizedDistinct{} distinct programs per attempt against \ResampleDistinct{}
for resampling. A capability probe down to 0.6B cannot carry the location comparison at all, because
\LadderUnparsed{} of the span arms' spliced programs do not parse at that scale, so we restrict
every localization conclusion to the 24--32B models tested.
\end{abstract}

\section{Introduction}

Testing and debugging are explicitly part of verifiable coding: tests provide executable evidence
that a behavior is wrong, while fault localization proposes where to intervene. Large language
models (LLMs) make the intervention easy to generate, but they also introduce a basic evaluation
problem. If a model repairs a localized span, did the test signal identify useful code, did any
small edit have a better chance than regenerating the whole program, or did an additional sample
provide all of the benefit?

These explanations require different controls. Recent placebo-controlled studies of frozen small
code models find that blind resampling can match or outperform feedback-conditioned self-repair
\citep{iscanfalsification,iscanform,verma2026}. In parallel, fault-localization studies show that
more precise locations can improve repair when locations are supplied by developer patches or
other oracles \citep{sepidband2026,liu2026}. Neither result answers whether \emph{predicted},
test-derived localization makes a targeted retry preferable to a fresh attempt. Refining the whole
program has been argued to beat localized hunks \citep{xu2024}, and the closest infilling work
varies masked locations \citep{koutcheme2023}, but neither separates the fresh sample from the edit
size with a matched attempt budget and a random-location control.

We test this missing comparison on the same initially failing candidate with three arms: blind
whole-solution resampling; Ochiai spectrum-based fault localization (SBFL) from public tests,
followed by fill-in-the-middle (FIM) replacement of the most suspicious tied span; and a disjoint,
same-length random-code-span placebo. This design separates two questions:
\begin{enumerate}
  \item Does the chosen location add value over editing an equally sized but unrelated code region?
  \item Does localized editing add value over spending the same number of attempts on fresh solutions?
\end{enumerate}

We declared a demanding success rule before running the experiment, recorded in the project's
kickoff documentation: localized infilling had to beat both controls, exceed the placebo by at
least five percentage points with Holm-adjusted $p<.05$, and replicate across at least two model
families. We call this rule \emph{pre-specified} rather than \emph{preregistered} because the study
plan was not registered with an external service. The rule was not met: the placebo margin is
$\PlaceboDiff$ points rather than five, no per-model test
of the placebo comparison survives correction under either analysis, and localized infilling loses
to blind resampling instead of beating it. Its replication clause is satisfied only in that last,
inverted sense. We report what the evidence does support instead.

What it supports is a four-part answer. Localization is usually unavailable: only
\PublicLocalizableRate{} of failing candidates expose a failing public test with a usable spectrum.
Where it is available, a fresh attempt beats a targeted one at a matched attempt count
(\ResampleDiscordance{}, $p={\ResampleP}$), resolving in two of the three original checkpoints
under both analyses and again in a fourth model from a third family (\MistralDiff{} points, 95\% CI
\MistralCI{}), so the cross-family bar we set in advance is met by the method losing to its own
control. Localized infilling does lead an unrelated edit of the same size when pairs are pooled
(\PlaceboDiscordance{}, Holm $p={\HolmPlacebo}$), but in no individual model under the primary
analysis, so we carry it as suggestive. Widening the edit does not rescue it, and re-pricing
attempts as tokens narrows rather than overturns the result: localization leads only below
\CrossoverTokens{} tokens, the price of one fresh attempt.

Our contribution is evaluative rather than a new repair algorithm. We provide (1) a two-control
protocol for test-guided repair, in which the random-span placebo is what separates ``the location
was right'' from ``the edit was small''; (2) a localizability audit over \DeadCandidates{} failures reporting how often public tests
support localization at all; (3) a budget analysis in both attempts and tokens, with the
edit-diversity measurements that explain why the two disagree; and (4) five concrete safeguards
(Appendix~\ref{app:defects}), each of which silently changed a measured result in our own pipeline
before it was fixed. We make no formal-verification or correctness-guarantee claim.

\section{Related work}

\paragraph{Feedback and blind resampling.}
\citet{iscanfalsification,iscanform} use preregistered placebo-controlled decompositions of
self-repair in frozen small code models. \citet{verma2026} finds blind retry to be a particularly
strong baseline below 7B parameters. These studies regenerate an entire solution. We retain their
blind control but ask whether restricting the edit to a test-localized span changes the comparison
at larger model scale.

\paragraph{Localization for LLM repair.}
Locations in this literature come from oracles or from predictors. On the oracle side, using 500
SWE-bench Verified instances and ground-truth patch locations, \citet{sepidband2026} find that file
localization is crucial and that adding line context can help or hurt depending on how it is
combined with broader context, and \citet{townsend2026} feed ground-truth locations to Agentless at
three granularities on SWE-Bench-Mini, finding function level highest but task dependent. On the
predictor side, SHERLOC performs structured, agentic repository localization and measures downstream
repair \citep{tamoyan2026}, \citet{sepidband2026rgfl} make the localizer reason over a repository
exceeding the context window, and \citet{alawad2026} runs the closest contrast to ours at file
granularity, where pooled resolved rate rises from 44.7\% to 48.9--49.1\% with predicted locations
and to 52.4\% with gold ones. None of these, so far as we could verify, reports an
untargeted-regeneration arm at a matched attempt budget or a random-location control, so
localization gains and retry-alone gains remain unseparated in the published record; and the setting
is complementary rather than overlapping, file-level selection inside a repository against
single-function repair, where regenerating everything is the control that matters most.
\citet{liu2026} vary localization precision across two APR techniques and three LLMs, and report
that imprecise fault localization makes repair harder and widens the gap between techniques.
\citet{fan2023} match our setting most closely, repairing failing Codex solutions under unguided,
line-guided and statement-guided Codex editing and finding guided repair ``nearly comparable'' to
unguided; their unguided arm still edits the failed program rather than drawing a fresh sample, and
they run no random-location control.

\paragraph{Targeted intervention against re-running.}
\citet{luan2026} ask our question in a different unit, whether multi-agent repair methods
``causally repair \dots{} or merely stochastically repair by leveraging the randomness of LLM
sampling''. Replaying a trajectory to an anchor and regenerating downstream, they find symptom-driven
intervention repairs 20.15\% of 536 annotated failures against 6.90\% for unguided rerunning, the
opposite sign to ours. Their anchor replays a prefix reconstructed from recorded logs and still
regenerates a whole downstream trajectory, where our span keeps a fixed prefix and suffix around a
gap of a few tokens. \citet{pdb2026} measure the complementary thing, that frontier models ``often
regenerate correct but over-edited solutions'', proposing edit-level precision; our size-matched
placebo is the control-side analogue.

\paragraph{Infilling and patch structure.}
\citet{xia2022} study pre-trained models for repair at scale, reporting that suffix context, which
an infilling formulation adds, yields more fixes at a higher compilation rate. Their complete-function setting ``makes no assumption of \dots{} the location of
the bug'' while their infilling and single-line settings are given it; it counts patches per setting
rather than at a matched attempt budget, with no random-location arm and 2022-era models. Our no-op measurement is a possible cost of
that suffix conditioning, which our design does not isolate. \citet{xu2024} argue our result's direction as a
design principle: restricting an LLM to statement-level hunks ``hinders LLMs from exploring
potential patches beyond the given locations''. Theirs is a prompting method with no
random-location arm and no matched attempt budget, so it motivates our question but does not
answer it.
\citet{koutcheme2023} evaluate InCoder by masking up to 50 candidate spans and sampling repairs,
demonstrating that generative infilling is a viable repair primitive.
The closest published contrast in representation space is \citet{silva2025}, who fine-tune LoRA
adapters and compare a whole-function input against inputs carrying fault-localization comments or
an infilling mask, and a full fixed function output against the fixed chunk alone. They ``only
assume that one is able to identify a code region that encompasses all the buggy lines'', varying
how a supplied location is presented to a \emph{fine-tuned} model; we vary whether a test-derived
location is used at all, for frozen models, against a fresh sample at a matched attempt count and a
random-location control. Ochiai SBFL is already combined with LLM repair:
\citet{farzandway2025} rank lines by Ochiai score into a chain-of-thought loop on Codeflaws, passing
localization as prompt context rather than using it to constrain the edit region. Multi-hunk defects become
harder as patch divergence and dispersion grow \citep{nashid2025}, so our single contiguous-span
setting is a deliberately limited first test. Classic SBFL assigns line suspiciousness from
coverage spectra \citep{abreu2007}; we use Ochiai and avoid picking an arbitrary line when several
tie at the maximum.

\section{Experimental design}
\label{sec:method}

\subsection{Study population and localization}

For each model and benchmark, we first greedily generate one candidate per task. A \emph{dead}
candidate fails the full benchmark tests. We attempt localization only when at least one test in
the localizing suite fails and executing those tests yields line coverage. Thus all three-arm
analyses are conditional on the same localizable subset; resampling outcomes from nonlocalizable
dead candidates are never included.

We localize under two suites, because which failures a repair loop can even see turns out to
dominate everything else it does. The \emph{public} signal uses only the tests a deployed loop
has, namely HumanEval's and MBPP's original cases. The \emph{strong} signal additionally reads the
augmented suite (every base test plus up to 80 sampled augmented ones), which no deployed loop
has; we report it as an upper bound on available localization signal and never as a method. The
strong signal is also what makes the comparison affordable to power: it takes the localizable
population from \PublicLocalizable{} candidates to \StrongLocalizable{}. Because the three-arm
comparison conditions on localizable failures, we separately replay this decision for every dead
candidate and record why it succeeds or fails (Figure~\ref{fig:funnel}).

For source line $\ell$, Ochiai suspiciousness is
\begin{equation}
s(\ell)=\frac{n_f(\ell)}{\sqrt{N_f\,[n_f(\ell)+n_p(\ell)]}},
\end{equation}
where $N_f$ is the number of failing tests in the localizing suite and $n_f(\ell)$ and
$n_p(\ell)$ count failing and passing tests that execute $\ell$. We take the contiguous range
spanning all lines tied at the maximum score. Candidates are excluded if this range exceeds 60\% of
source lines or if no disjoint same-length code-only placebo span exists. Under the public signal
the augmented tests are used only for final scoring; under the strong signal they also inform the
spectrum, which is exactly why that condition is a bound and not a method.

\subsection{Three arms}

Every arm receives $K=\SampleBudget$ stochastic samples with temperature 0.8 and nucleus probability 0.95.
The arms differ as follows.
\begin{itemize}
  \item \textbf{Blind resample:} generate a fresh complete solution from the original problem,
  without the failed candidate, coverage, or test result.
  \item \textbf{Localized infill:} replace the maximum-suspiciousness span while conditioning on
  its unchanged prefix and suffix, using native FIM tokens where present and a marked gap in a chat
  prompt otherwise. Decoding stops at the model's own FIM terminators, read from its tokenizer
  rather than assumed, and the completion is cut back to the middle it was asked for, since an
  instruct model handed a raw FIM prompt keeps going into a test harness and splicing that overrun
  makes the file unparseable. The cuts are structural (a markdown fence, a dedent out of the span's
  block, or a verbatim suffix) and identical for this arm and the placebo, so they cannot favour
  either (\S\ref{sec:threats}).
  \item \textbf{Random-span placebo:} replace a disjoint span of the same line length, sampled only
  from executable code lines rather than comments, docstrings, or blanks. It uses the same native
  or prompted infilling path as the localized arm.
\end{itemize}

The budget is matched in \emph{number of attempts}. Span arms allow 256 new tokens and
whole-solution resampling 512 on
HumanEval+ and MBPP+ and 1536 on LiveCodeBench, whose tasks are whole classes rather than single
functions and which truncated badly at a lower ceiling. The ceilings are far above what either arm
typically uses on its own benchmark, so they are not what drives the token gap we report; the arms'
measured output lengths are. Because attempts and tokens are different currencies, we report
the comparison in both, and treat neither as the definitive one.

\subsection{Models, benchmarks, and scoring}

We evaluate Qwen2.5-Coder-32B-Instruct-AWQ, Qwen3.6-27B and Gemma-4-26B-A4B-it, and add
Mistral-Small-3.2-24B-Instruct as a cross-family replication. The first uses 4-bit AWQ weights, the
others local unquantized checkpoints. Both Qwen tokenizers expose FIM tokens; Gemma and Mistral use
prompted pseudo-FIM. Mistral is a declared independent replication with the same estimator rather
than a fourth cell of the three-model analysis: adding a model to a correction family after that
family has been analysed and its plan published would change the multiplicity structure of an
already-reported result. It ran on HumanEval+ and MBPP+ only, so its \MistralPairs{} candidates are
not part of the \DeadCandidates{} the three-model analysis draws on. Thinking is
disabled through Qwen3.6's chat-template option so a reasoning preamble does not consume the code
budget. All weights are frozen; there is no training or adaptation.

The evaluated models are the object of study rather than tools used to conduct it. Separately, and
disclosed under the workshop's LLM policy: the harness, analysis scripts, and manuscript text were
written with substantial AI coding assistance under author direction. The experimental design, the
pre-specified decision rule, the choice of controls, and every claim in this paper are the authors'
responsibility, and each reported number is regenerated from the result files by a checked script
rather than transcribed.

HumanEval+ v0.1.10 and MBPP+ v0.2.0 supply both the base tests and the augmented ones
\citep{liu2023evalplus}. A third benchmark, the function-call subset of LiveCodeBench
\citep{jain2024} filtered to problems dated on or after 2024-07-01, contributes harder and more
recent problems; we do not call it contamination-free, because we did not verify any checkpoint's
training cutoff, and \S\ref{sec:threats} reports why that caveat is load-bearing there. An arm
\emph{unlocks} a candidate--task pair if any of its $\SampleBudget$ completions passes every
available test. Each EvalPlus test invocation is bounded by EvalPlus's own rule, the larger of one
second and four times the reference solution's runtime on that input, rather than by a flat constant
(\S\ref{sec:threats}); LiveCodeBench, which has no reference timing, uses a flat five-second bound.
Our scorer adopts that timeout rule but compares outputs to the reference by exact equality rather
than reusing EvalPlus's per-task tolerance and order-insensitivity logic, so our absolute rates are
not interchangeable with published EvalPlus numbers; every arm is scored identically, so it cannot
favour an arm. On LiveCodeBench a baseline generation that hit the token ceiling is excluded from
the dead population rather than counted as a failure, since truncation is a harness outcome and not
a model one; that removes \TruncExcludedCoder{}, \TruncExcludedQwen{} and \TruncExcludedGemma{}
candidates respectively, which is why the dead counts below are smaller than the raw pass rates in
\S\ref{sec:threats} imply.
Experiments ran on one NVIDIA DGX Spark with 128GB unified memory; generated-token counts are
measured per attempt, but wall-clock durations were not systematically recorded.

\paragraph{A capability probe, not a fourth model.}
Because every arm conditions on failures, a model's worth to this design is its failure count, not
its benchmark score, and that count falls sharply as capability rises: on LiveCodeBench the three
models yield \LcbDeadByModel{} dead candidates in order of increasing pass rate. We therefore
probed \emph{downward}, on Qwen3-0.6B, 1.7B and 4B over HumanEval+ and MBPP+
(Appendix~\ref{app:ladder}).

\subsection{Inference}

The kickoff decision rule (Appendix~\ref{app:prereg}) is written in terms of \emph{unlocks}:
whether any of $K$ samples passes. The Wave 1B analysis plan, specified before its generations,
made the \emph{per-attempt} success rate primary and the unlock test secondary, on the ground that
collapsing $\SampleBudget$ samples to one bit discards most of the data. We report both outcomes and
label their distinct roles.

Both analyses are per model (Table~\ref{tab:primary}), pooling each model's tasks across
benchmarks and Holm-correcting the three tests within each comparison family. The primary one takes
the difference in per-attempt success rate with a task-clustered bootstrap: a task's
$\SampleBudget$ samples are correlated, so tasks are the resampled unit and both arms are drawn
together to preserve the pairing. The secondary one uses a two-sided exact McNemar/binomial test on
discordant unlock outcomes. Pooled task--model-pair tests (Table~\ref{tab:comparisons}) are
sensitivity summaries under both plans: different models are not exchangeable replicates. We also
report each arm's unlock@$k$ curve under the standard unbiased estimator. All use the same
generations, and the released outcome files retain every attempt rather than only one bit per task.

All three arms come from a single pass of one harness on the same candidates, so no comparison here
splices runs made under different decoding rules.

\section{Results}
\label{sec:results}

\subsection{How often can test-guided repair fire at all?}

\begin{figure}[t]
\centering
\includegraphics[width=.80\linewidth]{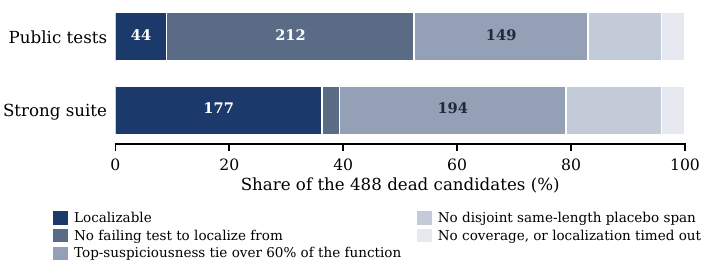}
\caption{Where \DeadCandidates{} dead candidates go (three models across HumanEval+, MBPP+ and
LiveCodeBench). The dark segment is the localizable population the span arms can run on. Under
public tests the dominant loss is an absent signal; under the strong suite, which no deployed loop
has, it becomes a tied top-suspiciousness block too diffuse to edit.}
\label{fig:funnel}
\end{figure}

Figure~\ref{fig:funnel} replays the localization decision on all \DeadCandidates{} dead candidates.
Public tests support localization for \PublicLocalizable{} of them (\PublicLocalizableRate{}). The
dominant loss is not a weak localizer but an absent signal: \NoFailingPublicTest{} candidates pass
every public test and fail only hidden ones, so no spectrum exists to compute. Localizing from the
augmented suite instead raises coverage to \StrongLocalizable{} (\StrongLocalizableRate{}) and
nearly eliminates that category, at which point the binding constraint becomes structure rather
than tests: the maximum-suspiciousness tie spans more than 60\% of the function for most of the
rest, because straight-line statements co-execute under every test. Where localization does apply
the span is small: a median of \SpanMedianPublic{} of the file under the public signal and
\SpanMedianStrong{} under the strong one.

Two consequences matter beyond this study. On this population a deployed repair loop that waits for
a visible failing test could act on only \PublicLocalizableRate{} of failures, and the ones it does
see are a biased sample: the strong signal adds \StrongOnlyLocalizable{} candidates whose bugs
survive every public test. What bounds test-guided repair here is the strength of the available
verifier rather than the quality of the localizer.

\subsection{Does the location matter, and does it beat starting over?}

Table~\ref{tab:primary} gives the primary per-model analysis and Table~\ref{tab:comparisons} the
pooled sensitivity summary. Take the placebo comparison first, because it is the one the placebo
exists to make: does an edit at the implicated location do something an equally sized edit at an
unrelated one does not?

Under the primary analysis it does not resolve. The per-attempt difference is positive in every
model, but it \PrimaryPlaceboVerdict{} after Holm correction (best Holm
$p={\PrimaryPlaceboBestHolm}$), and the secondary per-model unlock test resolves in
none of the three either. Only the pooled summary crosses the threshold: localized
infilling unlocks tasks the placebo does not far more often than the reverse
(\PlaceboDiscordance{}, exact $p={\PlaceboP}$, Holm-adjusted $p={\HolmPlacebo}$), at
\LocalizedAttemptRate{} of attempts against \PlaceboAttemptRate{}, a difference of $\PlaceboDiff$
points (95\% CI \PlaceboCI{}). That pooling treats three models as exchangeable replicates, which
both of our plans decline to do for a confirmatory claim, so the honest statement is that the
location plausibly matters and this experiment does not establish it.

Against blind resampling at a matched number of attempts the ordering reverses, the margin is far
larger (\ResampleDiscordance{}, $p={\ResampleP}$; \ResampleAttemptRate{} of blind attempts succeed
against \LocalizedAttemptRateVsResample{} of localized ones), and this is the one result that resolves within
individual models under \emph{both} analyses. The primary attempt-level test survives Holm correction in
\PrimaryResampleModels{} of the three checkpoints, and the secondary unlock test survives it in the
same two (Table~\ref{tab:primary}); the third resolves under neither, on too few discordant pairs to
separate a null from a small effect. Our pre-specified rule demanded
replication across at least two model \emph{families}. Both resolving checkpoints are Qwen, which
alone would not clear that bar, so we ran a fourth model from a third family. Mistral-Small-24B
reproduces the effect on \MistralPairs{} candidates at \MistralDiff{} points
(\MistralLocalizedRate{} of localized attempts succeed against \MistralResampleRate{} of blind
ones, 95\% CI \MistralCI{}), the largest margin in the study; its unlock discordance is
\MistralDisc{} ($p={\MistralMcNemarP}$), meaning resampling unlocked twenty-five candidates
localization did not and localization unlocked none that resampling did not. The bar is therefore
met, by the negative result. Table~\ref{tab:primary} collects all four.

One caveat we checked rather than waved through. Mistral's spliced programs fail to parse in
\MistralUnparseable{} of attempts against \FinalUnparseable{} for the others, so part of that
margin could be an interface failure rather than a localization failure. Restricted to attempts
that do parse, the difference is \MistralParseDiff{} points (95\% CI \MistralParseCI{}): it
widens rather than narrows, so it is not an artifact of malformed infills. Mistral also regenerates
the removed span verbatim in only \MistralNoop{} of attempts, about half the pooled rate of the
others, so it explores more and still loses by more. Against its own placebo it behaves like every
other model, \MistralPlaceboDiff{} points ($p={\MistralPlaceboP}$), which does not resolve.

The one model that does not reproduce the effect, Gemma-4-26B-A4B, is both the only non-Qwen of the
original three and the only mixture-of-experts model on the roster; the three dense models all
reproduce it. With a single MoE model we cannot separate those explanations and do not try to.

We report the deployable condition separately and it does not resolve: restricted to the
\PublicLocalizable{} candidates localizable from public tests alone, the placebo comparison is
\PublicPlaceboDiscordance{} ($p={\PublicPlaceboP}$) and the resampling comparison
\PublicResampleDiscordance{} ($p={\PublicResampleP}$), neither surviving correction. Every
claim that resolves in this study therefore depends on a localization signal a deployed loop would
not have.

\begin{table}[t]
\caption{Per-model tests on the strong-signal population under both designated analyses. Primary
(Wave 1B plan, shipped in the supplement): per-attempt success rate, task-clustered bootstrap.
Secondary (kickoff rule): the unlock outcome, exact McNemar, discordance
localized-only\,:\,comparator-only. Holm correction runs across the three models of the
pre-specified family within each comparison and analysis; bold marks Holm $p<.05$. $\Delta$pp is
localized minus comparator. $^\dagger$Mistral-Small-24B is a declared independent replication with
the same estimator, so it carries no adjusted $p$: it is not a member of that correction family.}
\label{tab:primary}
\centering
\small
\input{tables/primary}
\end{table}

Halving the budget leaves both effects in place: re-run on the first \HalfBudget{} samples of the
same generations, which varies the budget and nothing else, the placebo discordance is
\HalfPlaceboDiscordance{} ($p={\HalfPlaceboP}$, $\HalfPlaceboDiff$ points against $\PlaceboDiff$ at
full budget) and the resampling discordance \HalfResampleDiscordance{} ($p={\HalfResampleP}$).
Because both budgets read the same generations this is a sensitivity check showing directional
stability, not an independent replication.

\subsection{Attempts are the wrong unit of cost}

\begin{figure}[t]
\centering
\includegraphics[width=.58\linewidth]{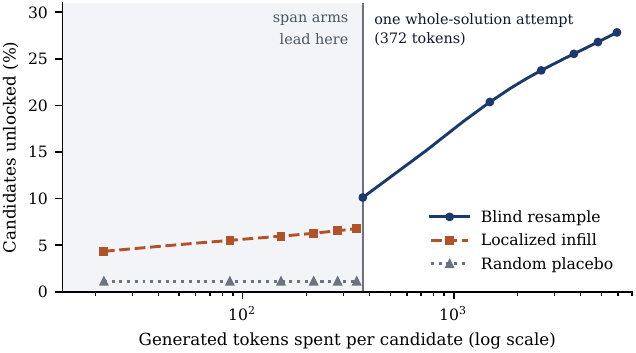}
\caption{Unlock rate against the generated tokens it costs, strong-signal population, $k=1$ to
$\SampleBudget$ attempts per arm priced at each arm's measured mean token cost. Localized infilling
leads only left of the line, where the blind arm cannot afford a single attempt; past it the blind
arm's first attempt already exceeds what \SampleBudget{} localized attempts reach in total, and the
span arms are flat.}
\label{fig:frontier}
\end{figure}

The arms do not spend the same compute per attempt: a span replacement averages \SpanTokens{}
generated tokens against \ResampleTokens{} for a whole solution, a factor of \TokenRatio{}. Per
attempt, blind resampling overtakes localized infilling by the third sample and keeps climbing.
Re-pricing in tokens (Figure~\ref{fig:frontier}) moves the crossover but does not remove it. Below
\CrossoverTokens{} generated tokens, the cost of one whole-solution attempt, the blind arm cannot buy a
single attempt while the span arm has already reached \SpanBelowCrossover{}; at and above that
price the blind arm leads at every budget we can evaluate, its first attempt reaching
\BlindFirstAttempt{} against the \SpanCeiling{} that \SampleBudget{} localized attempts reach in
total, and \BlindCeiling{} by the last. This normalizes by each arm's measured mean token cost
rather than capping tokens directly, so it ignores prompt and prefill cost, which favours the span
arms whose prompt carries the whole file. It also pools cells whose costs differ by an order of
magnitude: mean resampling length runs from \ResampleTokensMin{} tokens on the cheapest
model--benchmark cell to \ResampleTokensMax{} on the dearest, the latter being whole-class
LiveCodeBench tasks. \CrossoverTokens{} is therefore a pooled threshold for this population and
not a constant a deployment can plan against; the ordering it summarises is robust, the number is
not. A directly token-capped, wall-clock-matched comparison is the right way to settle it and we did
not run one.

So the honest reading is narrow. Localization is what a budget too small for one fresh attempt can
afford, which is a real regime for a latency-bound or per-token-billed deployment, and it is not an
efficiency advantage that survives into any larger budget. What both views agree on is that the
placebo never moves: extra budget spent at an unrelated location buys nothing at all.

\subsection{Why the span arms saturate, and whether more room helps}
\label{sec:saturation}

Infilling conditioned on a fixed prefix and suffix frequently regenerates the removed span character
for character: \LocalizedNoop{} of localized attempts and \PlaceboNoop{} of placebo attempts leave
the program unchanged and cannot unlock anything, against \ResampleNoop{} for blind resampling.
Distinct programs per attempt say the same thing: resampling \ResampleDistinct{}, the placebo
\PlaceboDistinct{}, localized infilling fewest at \LocalizedDistinct{}
(Table~\ref{tab:conditional}). This is the infilling analogue of the anchoring reported for
feedback-conditioned repair \citep{iscanfalsification,verma2026}, with the surrounding code as the
anchor rather than the failure message, and it binds hardest in the \emph{localized} arm: the
maximum-suspiciousness span is by construction the one the model just committed to and that its
prefix and suffix most strongly imply, so it is the span it is most likely to write again.

The wins are concentrated too, which is why the placebo line in Figure~\ref{fig:frontier} is flat
rather than shallow. Counting a task as saturated when an arm unlocks it on all but at most one of
its \SampleBudget{} attempts, the placebo saturates \PlaceboSaturated{} of the \PlaceboUnlocked{}
tasks it unlocks at all, localized infilling \LocalizedSaturated{} of \LocalizedUnlocked{}, and
resampling \ResampleSaturated{} of \ResampleUnlocked{}. The span arms' wins are the cases where the
replacement is already determined, so extra samples have nothing to find.

That invites the reading that the negative result is an artifact of degenerate attempts. It is not.
Scoring only attempts that changed the program narrows the gap without closing it
(\CondLocalized{} against blind resampling's \CondResample{}, the placebo \CondPlacebo{}), and
rejecting repeats does not rescue it either: collapsing each task's attempts to its
\DedupUnique{} distinct middles lifts the localized rate only to \DedupRate{}. The limit is the
proposal distribution, not the duplicates drawn from it.

So we changed the proposal distribution. On \WidenPairs{} candidates where both arms stay valid at
a larger size we widened the edit by two lines each side, moving the median span from
\WidenSpanFrom{} line to \WidenSpanTo{} and cutting verbatim regeneration to under five per cent.
Three variants ran, enumerated in Appendix~\ref{app:widening}, and none of them
rescues the method (Table~\ref{tab:widening}). The adaptive policy does not beat static narrow
editing in either correction-family model, and where it moves at all the movement is width rather
than adaptivity: Qwen3.6's $\WidenWidthGain$ points over narrow editing is matched exactly by
always editing wider, and adaptivity adds at most \AdaptiveOverWideMax{} points on top. The widened
arm still loses to blind resampling.

A capability probe running the same arms on Qwen3-0.6B, 1.7B and 4B produces \LadderPairs{}
localizable pairs and no usable \emph{location} comparison: \LadderUnparsed{} of those models'
spliced programs do not parse. Appendix~\ref{app:ladder} reports why that is a boundary on the
technique rather than a capability moderator, and what it does still establish.

\section{Implementation findings and threats to validity}
\label{sec:threats}

Five defects in our own pipeline each changed a measured result before it was fixed, and each is a
general risk for this kind of study rather than an artifact of our setup. Appendix~\ref{app:defects}
catalogues them: an unconstrained placebo span that
could land entirely in a docstring, a single-line \texttt{argmax} selecting by ordering rather than
evidence, a flat per-test timeout indefensible in both directions, an unregistered FIM terminator
that spliced unparseable text into \PreStopTokenUnparseable{} of localized attempts, and a
closed-fence extractor that silently returned whole truncated completions. They do not act in one
direction: the placebo defect weakened a control, the span-selection and terminator defects
weakened span arms, and the timeout and extractor defects changed which candidates were scored.

\paragraph{Multiplicity and what the tests carry.}
Holm correction across the \HolmFamily{} pooled comparisons leaves the resampling result at
$p={\HolmResample}$ and the placebo result at $p={\HolmPlacebo}$. Both of our plans locate the
confirmatory analysis per model rather than pooled, and there only the resampling comparison
resolves, in two of three checkpoints, under the primary attempt-level test and the
secondary unlock test alike (Table~\ref{tab:primary}). The placebo result is therefore pooled evidence with an interval
attached and not a confirmatory finding; the replication clause is met by Mistral reproducing the
loss; and the deployable public-test condition resolves nothing at all.

\paragraph{External validity.}
The analysis conditions on localizable failures from one greedy baseline per model, so it does not
describe parse failures, hidden-only failures, or diffuse spectra, and all tasks are function or
class level Python rather than repository repair or multi-hunk editing. Every model here without
native FIM tokens is also non-Qwen, so we measured the interface rather than assuming it: re-running
both Qwen models through the prompted path Gemma uses, on the same candidates and spans
(Appendix~\ref{app:pfim}), leaves the resampling loss negative in both (\PfimCoderDiff{} and
\PfimQwenDiff{} points), resolving in \PfimPromptedResolving{} of the two rather than
\PfimNativeResolving{}. Prompt format is therefore not what explains Gemma, which is also the only
mixture-of-experts checkpoint; nothing here separates family from architecture. The strong signal
reads augmented tests a deployed loop would not have, and every result that resolves depends on it.
On LiveCodeBench the pass-rate ordering tracks checkpoint recency rather than model size, which is
what contamination would look like on a pool whose cutoffs we did not verify, so we claim none.
Finally, test passage is bounded by the supplied suites and is not a proof of correctness.

\section{Broader impact}

A controlled baseline can stop organizations deploying a complicated repair loop where a fresh
attempt does as well, cutting engineering and inference cost. Conversely, test-guided
generation can produce plausible but incorrect or insecure code when suites are weak, and
localization may add unwarranted confidence by making a patch look surgical; generated patches
should remain sandboxed, reviewed, and tested against independent oracles. This work releases no
new model or sensitive dataset.

\section{Reproducibility}
\label{sec:repro}

The workshop supplement holds the source, unit tests, aggregate results and spans used here. Every table, figure and headline
quantity is emitted from the shipped result files rather than transcribed, the README maps each to
its file, and the build reruns every documented command in a fresh extraction while tracing the
emitter's file reads, so a missing input fails the build.
The paper build also rejects bare numerals and spelled-out magnitudes in prose, undefined
references, unincluded fragments, and designations that disagree with the public analysis record.
Weights and benchmarks are not redistributed; locks and durations were not preserved.

\bibliographystyle{plainnat}

\appendix
\section{The capability probe}
\label{app:ladder}

Running the same three arms on Qwen3-0.6B, 1.7B and 4B yields \LadderPairs{} localizable pairs,
more than the three deployment-scale models produce together, and no usable comparison between the
two span arms. Blind
resampling unlocks \LadderResample{} of \LadderResampleTasks{} tasks while localized infilling
unlocks \LadderLocalized{} and the placebo \LadderPlacebo{}. The reason is not that the location
stops mattering: \LadderUnparsed{} of these models' spliced programs do not parse at all, against
\FinalUnparseable{} for the deployment-scale models. \LadderParsed{} of the small models' splices are valid
Python, so the arm is not wholly degenerate, but it is not valid often enough for the placebo
contrast to carry a comparison.

We report this as a boundary on the technique and explicitly not as a capability moderator. A
moderator claim would need the small models' infills to be valid and merely unhelpful; here they
are invalid, and the two explanations (that sub-4B models cannot use a localized gap, and that they
cannot follow the raw fill-in-the-middle format at all) are not separated by this experiment.
Gemma-4 already uses a prompted pseudo-FIM path for want of native FIM tokens (\S\ref{sec:method}),
and whether that path rescues small models is the experiment we did not run. What the probe does
establish is that the blind-resampling result extends below 7B with a large margin, consistent with
the small-model findings we build on \citep{iscanfalsification,verma2026}.

\section{Implementation defects}
\label{app:defects}

\paragraph{Controls and localization.}
An unconstrained random span sometimes selected only a docstring or comment, which cannot change
execution and is a structurally weak placebo; the final sampler masks comments, docstrings and
blanks by tokenization and requires a disjoint span of identical line length. Straight-line
statements that co-execute under every test receive identical spectra, so a single-line
\texttt{argmax} selects by ordering rather than evidence; the final procedure spans the full
contiguous maximum-score range and rejects it when too diffuse.

\paragraph{Untrusted code needs more than a clock.}
A flat five-second per-test bound proved indefensible in both directions: a correct HumanEval/39
solution needs about 5.1 seconds on our hardware, so the same generation passed or failed with
machine load. EvalPlus scoring now follows EvalPlus's own rule, the larger of one second and four
times the reference solution's runtime on that input. A timeout also does not bound memory: one
candidate grew to 92GB resident and its worker was killed, deadlocking a shared process pool. Each
scoring job now runs in its own supervised process with an address-space limit, so a hang or an
out-of-memory candidate costs one task rather than the run.

\paragraph{Infilling needs its stop token, and truncation needs detection.}
Two extraction defects each turn a harness failure into an apparent model failure. The FIM
terminator was not registered as a stop token and was stripped from the decoded text, so span
generations ran past the intended middle and spliced unparseable text into
\PreStopTokenUnparseable{} of localized attempts. Registering the terminators, read from each
tokenizer rather than assumed, was necessary but not sufficient, because an instruct model handed a
raw FIM prompt writes the replacement and then a demonstration harness. That left
\UntrimmedUnparseable{} of localized attempts unparseable, and the structural cuts of
\S\ref{sec:method} bring that to \FinalUnparseable{}. Separately, extracting a
fenced block with a closed-fence pattern silently returns the whole completion when generation was
truncated before the closing fence, which misclassified \TruncQwen{} of \TruncTotal{} (Qwen3.6-27B)
and \TruncGemma{} of \TruncTotal{} (Gemma-4-26B-A4B) LiveCodeBench generations as genuine failures.
Raising the ceiling and detecting truncation leaves \TruncExcludedQwen{} and \TruncExcludedGemma{}
generations still truncated in the reported run; those are dropped from the dead population rather
than scored (\S\ref{sec:method}), a population change a reader cannot infer from the pass rates.
Both defects depress a treatment arm silently.

\section{Widening the localized edit}
\label{app:widening}

Table~\ref{tab:widening} is the falsification attempt described in \S\ref{sec:saturation}, reported
in full. Three arms ran against the narrow localized span on the \WidenPairs{} candidates where
every variant stays valid at the larger size: always editing two lines wider on each side, an
adaptive policy that widens only the \WidenTriggerRate{} of attempts whose narrow output was a
no-op or a duplicate, and a placebo held to the same realized edit size and generation-call count.
Registering the static-wide arm in advance is what makes the table readable: Qwen3.6's
$\WidenWidthGain$ points over narrow editing reads as evidence for adaptive localization until the
static-wide column shows the same margin, at which point the movement is width alone and the
adaptive policy contributes at most $\AdaptiveOverWideMax$ points on top of it. The bottom row is
the one that matters for the paper's claim: every widened variant still loses to blind resampling.

\begin{table}[t]
\caption{Widening the localized edit: per-attempt difference in points on the \WidenPairs{}
candidates where both arms stay valid at the larger size. Bold marks Holm $p<.05$ within the
pre-specified correction family. Rows two and three are the decomposition: width helps Qwen3.6,
adaptivity adds nothing on top of width.}
\label{tab:widening}
\centering
\small
\input{tables/widening}
\end{table}

\section{Generation integrity and pooled sensitivity}
\label{app:integrity}

Two tables sit behind claims the main text states in one line each.
Table~\ref{tab:conditional} is the arm-level accounting: how many attempts each arm spent, how
often it returned the program it was given, how many distinct programs it produced, how often the
splice failed to parse, and its success rate both unconditionally and restricted to attempts that
changed something. The last column is the answer to the obvious objection that the negative result
is an artifact of degenerate attempts: restricting to attempts that changed the program narrows the
gap and does not close or reverse it. All three arms are measured on the same strong-signal
population, so the rates are directly comparable; the resampling row carries slightly fewer
attempts because one localizable candidate has no resampling arm.

Table~\ref{tab:comparisons} is the pooled sensitivity summary. It is not the primary analysis and
is not offered as one: both of this project's planning documents locate the confirmatory test per
model, which is Table~\ref{tab:primary}. Its value is that it shows the two localization signals
side by side, and the public-signal rows are the deployable condition. Neither public-signal
comparison resolves, which is the result a reader planning a deployment should take from this
paper.

\begin{table}[t]
\caption{What each arm produced on the strong-signal population, and what it produced that could
count. Distinct/att.\ is unique spliced programs divided by attempts. Conditioning on a changed
program removes the no-op difference from the comparison and does not reverse it.}
\label{tab:conditional}
\centering
\small
\input{tables/conditional}
\end{table}

\begin{table}[t]
\caption{Both comparisons, both signals. Discordance is localized-only\,:\,comparator-only over
paired candidates; $p$ is a two-sided exact McNemar test on those pairs. Attempt-level rates treat
each of the \SampleBudget{} samples as an observation, with a task-clustered bootstrap interval on the
difference. Holm correction across these \HolmFamily{} tests leaves both strong-signal comparisons below $.05$
(resampling $p={\HolmResample}$, placebo $p={\HolmPlacebo}$) and neither public-signal comparison
below it. $N$ is one lower for the strong-signal resampling row because one localizable candidate
has no resampling arm. These pooled tests are sensitivity summaries under both of our analysis plans, not the primary
analysis; Table~\ref{tab:primary} is.}
\label{tab:comparisons}
\centering
\small
\input{tables/comparisons}
\end{table}

\section{Pre-specified decision rule and artifact map}
\label{app:prereg}

The decision rule below was recorded in the study plan before any results existed. Because the plan
was not registered with an external service, we call it pre-specified rather than preregistered. The
win bar required localized infilling to beat placebo by at least five percentage
points with Holm-adjusted $p<.05$, beat blind resampling, and replicate across at least two model
families. A localized-placebo tie or loss killed the localization mechanism claim; a failure to beat
resampling killed the practical recommendation. All three win clauses failed: the observed placebo
margin is $\PlaceboDiff$ points rather than five, no per-model placebo test survives correction
under either analysis, and localized infilling loses to blind resampling rather than beating it.
The replication clause is the only one satisfied, and only in the inverted direction, because what
replicates is the loss: it holds across two Qwen checkpoints and, in a later declared replication,
a third family. No confirmatory follow-on or positive-mechanism claim is therefore reported.

The workshop supplement ships \texttt{analysis\_plan.json}, a strict public record of the scientific
designations used here. It distinguishes the kickoff unlock rule from the Wave 1B primary outcome:
per-attempt success rate for localized versus size-matched placebo under the strong signal, using a
task-clustered bootstrap per model and Holm correction across models. The unlock-level test is
secondary. Its rationale, that collapsing $\SampleBudget$ samples to one bit discards most of the
data, is visible in this paper's own results. Section~\ref{sec:method} reports both outcomes, and
Table~\ref{tab:primary} reports the designated primary analysis.

The same public record specifies two additions: a Mistral cross-family replication and a re-run of
both Qwen models through the prompted pseudo-FIM path on the tasks they had already run natively.
Both are complete. Appendix~\ref{app:pfim} reports the interface ablation against its declared
interpretation rule, including the selected branch and the measured attenuation.

The workshop supplement maps Table~\ref{tab:comparisons} to the per-model and per-benchmark result
files behind it, and every reported figure to the statistics summary it is generated from. Generation checkpoints retain the exact
localized and placebo spans, which matters because placebo spans were seeded from
process-dependent hash values: they are recoverable from the released checkpoints but not
reconstructible from task identifiers alone, and the released harness derives seeds from SHA-256
instead. The localizable-task restriction is applied when loading every arm,
including resampling; this prevents totals over all dead tasks from being mixed with the smaller
paired population.

\section{The prompted pseudo-FIM interface ablation}
\label{app:pfim}

Design item (2) in the public analysis record is reported here regardless of its outcome. Every
model on this roster without native FIM tokens is also non-Qwen, so
family and interface are one partition and neither can be credited with Gemma's non-replication.
Forcing a Qwen down the prompted path varies the interface alone.

Both Qwen models were re-run on the same candidates, the same spans and the same funnel, differing
only in how the gap is presented: a chat request to fill a marked hole instead of raw FIM control
tokens. The blind-resampling arm is reused from the native run rather than redrawn. It never sees a
span and never uses FIM, so the interface cannot reach it, and holding the comparator fixed makes
any movement attributable to the interface; all \PfimPairs{} paired candidates carry the identical
resampling arm in both conditions. Both conditions in Table~\ref{tab:pfim} are recomputed over
these two models and this paired set under a two-model Holm family, so the native rows are not
Table~\ref{tab:primary}'s: comparing a two-test family against a three-test family would let a
change in multiplicity structure read as an interface effect.

\begin{table}[t]
\caption{The interface ablation. Both conditions, both Qwen models, the same \PfimPairs{} paired
candidates and the same comparator. Bold marks Holm $p<.05$ within the two-model family.}
\label{tab:pfim}
\centering
\small
\input{tables/pfim}
\end{table}

The prompted path is not degenerate here, which is what makes the comparison usable at all:
\PfimUnparseable{} of its localized splices fail to parse against \FinalUnparseable{} natively,
far from the \LadderUnparsed{} that voided the capability probe, and its no-op rate
(\PfimNoop{}) is essentially the native one. Measured on the localized arm alone, the interface
changes little: \PfimIfaceCoder{} points for Qwen2.5-Coder and \PfimIfaceQwen{} for Qwen3.6,
neither resolving.

We report the attenuation rather than only the branch. Qwen2.5-Coder's loss narrows enough to stop
resolving under correction, on the same candidates and the same $N$, so the interface is not
irrelevant to the size of the effect; what it does not do is reverse the sign in either model or
explain why Gemma behaves differently. The placebo comparison under the prompted path resolves in
one model, but that is one cell of a four-way exploration this ablation did not designate, and we
carry it as exploratory rather than as evidence for a location effect.

\end{document}

%% file: tables/macros.tex
\newcommand{\DeadCandidates}{488}
\newcommand{\PublicLocalizable}{44}
\newcommand{\PublicLocalizableRate}{9.0\%}
\newcommand{\StrongLocalizable}{177}
\newcommand{\StrongLocalizableRate}{36.3\%}
\newcommand{\NoFailingPublicTest}{212}
\newcommand{\StrongPairs}{177}
\newcommand{\PlaceboDiscordance}{11:1}
\newcommand{\PlaceboP}{.006}
\newcommand{\ResampleDiscordance}{3:40}
\newcommand{\ResampleP}{3.0\times 10^{-9}}
\newcommand{\LocalizedAttemptRate}{4.3\%}
\newcommand{\LocalizedAttemptRateVsResample}{4.4\%}
\newcommand{\PlaceboAttemptRate}{1.1\%}
\newcommand{\ResampleAttemptRate}{10.1\%}
\newcommand{\LocalizedUnlocked}{12}
\newcommand{\LocalizedSaturated}{6}
\newcommand{\PlaceboUnlocked}{2}
\newcommand{\PlaceboSaturated}{2}
\newcommand{\ResampleUnlocked}{49}
\newcommand{\ResampleSaturated}{2}
\newcommand{\PlaceboDiff}{+3.2}
\newcommand{\PlaceboCI}{[+0.5, +6.2]}
\newcommand{\PublicPlaceboDiscordance}{7:1}
\newcommand{\PublicPlaceboP}{.070}
\newcommand{\PublicResampleDiscordance}{2:10}
\newcommand{\PublicResampleP}{.039}
\newcommand{\LcbDeadByModel}{122, 46, 13}
\newcommand{\LadderPairs}{256}
\newcommand{\LadderUnparsed}{65.5\%}
\newcommand{\LadderParsed}{34.5\%}
\newcommand{\LadderLocalized}{1}
\newcommand{\LadderPlacebo}{0}
\newcommand{\LadderResample}{71}
\newcommand{\LadderResampleTasks}{263}
\newcommand{\CrossoverTokens}{372}
\newcommand{\SpanBelowCrossover}{6.8\%}
\newcommand{\SpanCeiling}{6.8\%}
\newcommand{\BlindFirstAttempt}{10.1\%}
\newcommand{\BlindCeiling}{27.8\%}
\newcommand{\SpanTokens}{21.7}
\newcommand{\ResampleTokens}{371.1}
\newcommand{\TokenRatio}{17.1}
\newcommand{\ResampleTokensMin}{119}
\newcommand{\ResampleTokensMax}{1213}
\newcommand{\TruncQwen}{78}
\newcommand{\TruncGemma}{70}
\newcommand{\TruncTotal}{160}
\newcommand{\PreStopTokenUnparseable}{13.4\%}
\newcommand{\UntrimmedUnparseable}{5.2\%}
\newcommand{\FinalUnparseable}{2.5\%}
\newcommand{\LocalizedNoop}{48.9\%}
\newcommand{\PlaceboNoop}{41.4\%}
\newcommand{\ResampleNoop}{10.2\%}
\newcommand{\PfimPairs}{142}
\newcommand{\PfimCoderDiff}{-4.9}
\newcommand{\PfimQwenDiff}{-9.6}
\newcommand{\PfimNativeResolving}{2}
\newcommand{\PfimPromptedResolving}{1}
\newcommand{\PfimUnparseable}{5.3\%}
\newcommand{\PfimNoop}{49.7\%}
\newcommand{\PfimIfaceCoder}{+2.3}
\newcommand{\PfimIfaceQwen}{-0.5}
\newcommand{\LocalizedDistinct}{0.23}
\newcommand{\ResampleDistinct}{0.83}
\newcommand{\StrongOnlyLocalizable}{62}
\newcommand{\HolmPlacebo}{.019}
\newcommand{\HolmResample}{1.2\times 10^{-8}}
\newcommand{\HolmFamily}{4}
\newcommand{\HalfBudget}{8}
\newcommand{\HalfPlaceboDiscordance}{10:1}
\newcommand{\HalfPlaceboP}{.012}
\newcommand{\HalfPlaceboDiff}{+3.1}
\newcommand{\HalfResampleDiscordance}{4:34}
\newcommand{\HalfResampleP}{6.0\times 10^{-7}}
\newcommand{\MistralPairs}{62}
\newcommand{\MistralLocalizedRate}{0.6\%}
\newcommand{\MistralResampleRate}{11.9\%}
\newcommand{\MistralDiff}{-11.3}
\newcommand{\MistralCI}{[-16.6, -6.8]}
\newcommand{\MistralDisc}{0:25}
\newcommand{\MistralMcNemarP}{6.0\times 10^{-8}}
\newcommand{\MistralParseDiff}{-12.2}
\newcommand{\MistralParseCI}{[-17.8, -7.2]}
\newcommand{\MistralUnparseable}{19.9\%}
\newcommand{\MistralNoop}{26.0\%}
\newcommand{\MistralPlaceboDiff}{+0.6}
\newcommand{\MistralPlaceboP}{.258}
\newcommand{\WidenPairs}{155}
\newcommand{\WidenTriggerRate}{80.6\%}
\newcommand{\WidenSpanFrom}{1}
\newcommand{\WidenSpanTo}{5}
\newcommand{\AdaptiveOverWideMax}{+0.6}
\newcommand{\WidenWidthGain}{+4.9}
\newcommand{\SampleBudget}{16}
\newcommand{\PrimaryPlaceboVerdict}{does not resolve in any model}
\newcommand{\PrimaryResampleModels}{2}
\newcommand{\PrimaryPlaceboBestHolm}{.087}
\newcommand{\CondLocalized}{8.5\%}
\newcommand{\CondResample}{11.3\%}
\newcommand{\CondPlacebo}{1.9\%}
\newcommand{\DedupUnique}{663}
\newcommand{\DedupRate}{5.6\%}
\newcommand{\PlaceboDistinct}{0.31}
\newcommand{\SpanMedianPublic}{0.07}
\newcommand{\SpanMedianStrong}{0.06}
\newcommand{\TruncExcludedQwen}{16}
\newcommand{\TruncExcludedGemma}{17}
\newcommand{\TruncExcludedCoder}{0}

%% file: tables/primary.tex
\begin{tabular}{lrrrrr}
\toprule
 & & \multicolumn{2}{c}{Primary: per attempt} & \multicolumn{2}{c}{Secondary: unlock} \\
Model & $N$ & $\Delta$pp (95\% CI) & Holm $p$ & Disc. & Holm $p$ \\
\midrule
\multicolumn{6}{l}{\emph{vs.\ random placebo}} \\
\quad Qwen2.5-Coder-32B & 91 & $+1.3$ $[+0.1, +3.5]$ & $.087$ & 4:0 & $.375$ \\
\quad Qwen3.6-27B & 52 & $+2.9$ $[-3.1, +9.2]$ & $.382$ & 4:1 & $.500$ \\
\quad Gemma-4-26B-A4B & 34 & $+8.8$ $[+0.0, +20.6]$ & $.156$ & 3:0 & $.500$ \\
\quad Mistral-Small-24B$^\dagger$ & 62 & $+0.6$ $[+0.0, +1.7]$ & --- & 2:0 & --- \\
\multicolumn{6}{l}{\emph{vs.\ blind resample}} \\
\quad Qwen2.5-Coder-32B & 91 & $-7.1$ $[-12.3, -2.3]$ & $\mathbf{.016}$ & 1:19 & $\mathbf{<.001}$ \\
\quad Qwen3.6-27B & 51 & $-9.1$ $[-16.4, -2.2]$ & $\mathbf{.022}$ & 1:19 & $\mathbf{<.001}$ \\
\quad Gemma-4-26B-A4B & 34 & $+2.9$ $[-2.2, +10.3]$ & $.435$ & 1:2 & $1.000$ \\
\quad Mistral-Small-24B$^\dagger$ & 62 & $-11.3$ $[-16.6, -6.8]$ & --- & 0:25 & --- \\
\bottomrule
\end{tabular}

%% file: tables/widening.tex
\begin{tabular}{lrrr}
\toprule
Comparison & Qwen2.5-Coder & Qwen3.6 & Gemma-4 \\
\midrule
Adaptive vs static narrow & $+0.5$ & $+4.9$ & $-6.2$ \\
Static wide vs static narrow & $+0.4$ & $+4.9$ & $-6.9$ \\
Adaptive vs static wide & $+0.1$ & $+0.0$ & $+0.6$ \\
Adaptive vs blind resample & $\mathbf{-6.1}$ & $-5.5$ & $-2.1$ \\
\bottomrule
\end{tabular}

%% file: tables/conditional.tex
\begin{tabular}{lrrrrrr}
\toprule
Arm & Attempts & No-op & Distinct/att. & Unparsed & Success/att. & Success $\mid$ changed \\
\midrule
Blind resample & 2815 & 10.2\% & 0.83 & 0.5\% & 10.12\% & 11.27\% \\
Localized infill & 2832 & 48.9\% & 0.23 & 2.5\% & 4.34\% & 8.51\% \\
Random placebo & 2832 & 41.4\% & 0.31 & 2.5\% & 1.13\% & 1.93\% \\
\bottomrule
\end{tabular}

%% file: tables/comparisons.tex
\begin{tabular}{llrrrrrrr}
\toprule
Signal & Comparator & $N$ & Disc. & $p$ & Localized & Comparator & $\Delta$pp & 95\% CI \\
\midrule
Public & Random placebo & 44 & 7:1 & $.070$ & 10.7\% & 4.5\% & $+6.1$ & $[-2.1, +14.9]$ \\
Public & Blind resample & 44 & 2:10 & $.039$ & 10.7\% & 12.4\% & $-1.7$ & $[-9.8, +7.0]$ \\
Strong & Random placebo & 177 & 11:1 & $.006$ & 4.3\% & 1.1\% & $+3.2$ & $[+0.5, +6.2]$ \\
Strong & Blind resample & 176 & 3:40 & $3.0\times 10^{-9}$ & 4.4\% & 10.1\% & $-5.8$ & $[-9.3, -2.2]$ \\
\bottomrule
\end{tabular}

%% file: tables/pfim.tex
\begin{tabular}{lrrrr}
\toprule
Condition / model & $N$ & Localized & $\Delta$pp vs resample (95\% CI) & Holm $p$ \\
\midrule
\multicolumn{5}{l}{\emph{Native FIM tokens}} \\
\quad Qwen2.5-Coder-32B & 91 & 2.40\% & $-7.1$ $[-12.2, -2.2]$ & $\mathbf{.011}$ \\
\quad Qwen3.6-27B & 51 & 4.90\% & $-9.1$ $[-16.3, -2.0]$ & $\mathbf{.015}$ \\
\multicolumn{5}{l}{\emph{Prompted pseudo-FIM}} \\
\quad Qwen2.5-Coder-32B & 91 & 4.67\% & $-4.9$ $[-9.9, +0.4]$ & $.072$ \\
\quad Qwen3.6-27B & 51 & 4.41\% & $-9.6$ $[-17.2, -1.7]$ & $\mathbf{.041}$ \\
\bottomrule
\end{tabular}